\documentclass[a4paper,11pt]{article}
\usepackage{amsmath}
\usepackage{amsfonts}
\usepackage{amssymb}
\usepackage{bm}
\usepackage[utf8x]{inputenc}
\usepackage{ae}
\usepackage{enumerate}
\usepackage[T1]{fontenc}
\usepackage[english]{babel}
\usepackage{braket}
\usepackage{graphicx}
\usepackage{makecell}
\usepackage{placeins}
\usepackage{tabularx}
\usepackage{tabulary}
\usepackage{setspace}
\usepackage{tablefootnote}
\usepackage[flushmargin]{footmisc}
\usepackage[comma]{natbib}
\usepackage{array}
\usepackage{float}
\usepackage{caption}
\usepackage{tikz}
\usetikzlibrary{shapes.geometric}
\usepackage[a4paper,left=2.5cm,right=2.5cm,top=3cm,bottom=3cm]{geometry}
\usepackage[colorinlistoftodos]{todonotes}
\usepackage{pdfpages}
\usepackage{longtable}
\usepackage{adjustbox}
\usepackage{booktabs}
\usepackage{multirow}
\usepackage{lscape}
\usepackage[colorlinks=true, allcolors=blue]{hyperref}
\usepackage{cleveref}
\usepackage[flushleft]{threeparttable}
\usepackage{inputenc}


\begin{document} \doublespacing \pagestyle{plain}
	
	\def\ci{\perp\!\!\!\perp}
	\begin{center}
		
		{\LARGE How to Use Data Science in Economics -- a Classroom Game Based on Cartel Detection}
		
		{\large \vspace{0.8cm}}
		
		{\large Hannes Wallimann and Silvio Sticher}\medskip

		{\small {University of Applied Sciences and Arts Lucerne, Institute of Tourism and Mobility} \bigskip }
		
		{\large \vspace{0.8cm}}
		
		{\large Version January 2024}\medskip
		
	\end{center}
	
	\smallskip

	\noindent \textbf{Abstract:} We present a classroom game that integrates economics and data-science competencies. In the first two parts of the game, participants assume the roles of firms in a procurement market, where they must either adopt competitive behaviors or have the option to engage in collusion. Success in these parts hinges on their comprehension of market dynamics. In the third part of the game, participants transition to the role of competition-authority members. Drawing from recent literature on machine-learning-based cartel detection, they analyze the bids for patterns indicative of collusive (cartel) behavior. In this part of the game, success depends on data-science skills. We offer a detailed discussion on implementing the game, emphasizing considerations for accommodating diverging levels of preexisting knowledge in data science.
	
	{\small \smallskip }
	{\small \smallskip }
	{\small \smallskip }
	
	
	{\small \smallskip }
	{\small \smallskip }
	{\small \smallskip }
	
	
	\bigskip
	\bigskip
	\bigskip
	\bigskip
	
	{\small {\scriptsize 
			\begin{spacing}{1.5}\noindent  
				\textbf{Addresses for correspondence:} Hannes Wallimann, University of Applied Sciences and Arts Lucerne, Rösslimatte 48, 6002 Lucerne, \href{mailto:hannes.wallimann@hslu.ch}{hannes.wallimann@hslu.ch}; Silvio Sticher, \href{mailto:silvio.sticher@hslu.ch}{silvio.sticher@hslu.ch}.
			\end{spacing}
			
		}\thispagestyle{empty}\pagebreak  }

	{\small \renewcommand{\thefootnote}{\arabic{footnote}} %
		\setcounter{footnote}{0}  \pagebreak \setcounter{footnote}{0} \pagebreak %
		\setcounter{page}{1} }
	
	\section{Introduction}\label{introduction}
	
	Ever since the advent of game theory in the mid-20th century and George Stigler's "The Theory of Oligopoly" \citep{stigler1964theory}, the study of cartels has been an integral component of any economics curriculum. Cartels, which cause prices to increase and/or the quality of products and services to decrease, not only intrigue researchers but are of utmost importance to policymakers, who have to make sure that public funds are used in an efficient manner. Moreover, cartel formation is illegal in most countries worldwide, and firms must expect high sanctions when a conspiracy is detected.
	
	In recent years, the erstwhile predominantly theoretic field of industrial organization has gradually shifted towards empirical research. One contributing factor is the increasing potential to harness large data sets for scrutinizing markets and identifying patterns indicative of cartel presence \citep[see, e.g., ][]{Harrington2008}, an indispensable first step in fighting cartels. According to the \cite{OECD2022}, in the past years, the literature on cartel screening has primarily focused on the utilization of machine learning. 
	
	From the students' and practitioners' point of view, such developments pose a challenge, as modern data-science skills are suddenly required in addition to understanding (imperfect) markets. The added complexity may hinder their ability to translate motivation into action, a gap that educationalists sometimes propose narrowing through the use of "edutainment" \citep[see, e.g., ][]{prensky2003digital}. As \cite{alsawaier2018effect} summarizes, this can be achieved either by adding game elements and mechanics to the educational environment ("gamification") or by simulating content using actual games ("game-based learning"). We follow the second approach by incorporating a game with defined learning outcomes \citep{plass2015foundations}, following the argument games motivate learners to stay engaged over a long period through a series of motivational game features.
	However, we also consider Folmar's cautionary note, who argues that classroom games carry the risk of short-lived engagement and motivation, ending with the game \citep{folmar2015game}. To address this, we break the game into several parts and integrate it with other teaching elements, such as theory inputs.
	
	The learning objectives of our classroom game are that participants gain an understanding of how data science is employed in cartel detection and that they are able to utilize corresponding tools accordingly. The virtual setting of our classroom game is the procurement procedure in the construction sector. In public-sector contracting, procurement agencies invite firms to submit bids for projects put to tender \citep[for Switzerland, see, e.g.,][]{wallimann2022machine}. Therefore, a procurement agency announces future contracts and deadlines for bid submissions. Moreover, the agency provides interested firms with relevant information for the contract. Then---if interested---construction firms prepare and submit their bids. After a pre-announced date, the procurement agency proceeds with a detailed examination of the firms' bids. When it comes to awarding contracts, an array of criteria come into play, encompassing references, quality, environmental considerations, and more. Among these, the pivotal criterion, especially in the field of road construction (and civil engineering in general), pertains to the offered price (or bid).	The thing to notice is that in some markets, there exist traits conducive to cartel formation \citep{harrington2006behavioral} whereby firms conspire to raise prices (or lower the quality of goods and services). Collusion is more likely with a small number of firms, little or no entry, repetitive bidding, and stable demand \citep[see, e.g., ][]{OECD2009}. As a result, the construction industry is particularly prone to collusion and is therefore chosen as our field of application.
	
	In the first two parts of our classroom game, participants take on the role of construction companies that can collude---or not---by submitting bids for projects put to tender. They can earn victory points as an analogy to the profits that firms obtain, both lawfully and illicitly. In the final part of the game, they take on the role of staff members within an antitrust authority applying machine learning algorithms in order to flag cartels. In this sequence, participants' victory points are modified depending on their success in applying the relevant data-science techniques, drawing an analogy to the recognition and career opportunities for antitrust professionals. Depending on the requirement level of the course or training sequence, this can involve sophisticated approaches, such as training models and selecting algorithms, or more accessible options, including purely visual analysis.
	
	The rest of our paper is organized as follows. In Section \ref{Sec:Literature}, we offer a concise summary of the relevant literature. In Section \ref{Sec:Classgame}, we specify the classroom game by describing the setting and parameters, along with detailing the game mechanics. This includes the bidding and awarding procedure of the first two parts of the game and the cartel-flagging mechanism of the final part. To support lecturers, we also provide an illustrative example on how to compute victory points. In Section \ref{Sec:Recomm}, we discuss the implementation of the classroom game in different settings with an emphasis on skill requirements. We conclude in Section \ref{Conc}, and provide supporting teaching material in the Appendix.
	
	\section{Related literature} \label{Sec:Literature}
	
	Our proposed classroom game relates both to recent advances in cartel detection as well as game-based learning in economics.
	
	As stated in the introduction, our game is situated within the evolving landscape of economists' tasks, particularly the increasing integration of machine-learning techniques for detecting cartel behavior. This topic has aroused the interest of economists in resent years \citep[see, e.g.,][]{varian2014big,athey2019machine}. Harrington's insights \citep{harrington2006behavioral}, which suggest that the construction sector is particularly susceptible to collusion, have been confirmed on several occasions \citep[see, e.g., ][for examples from Sweden, Switzerland, and Japan, respectively]{bergman2020interactions,imhof2018screening,ishii2014bid}. The use of statistical methods to screen markets to detect patterns of a cartel was proposed, among others, by \cite{Porter1993} and \cite{Harrington2008}. Following many machine-learning applications \citep[see, e.g., ][]{imhof2021detecting,harrington2022cartel,huber2019machine,rodriguez2022collusion,silveira2022won}, we also use behavioral screens as predictors in supervised models to detect collusive behavior.
	
	Game-based learning and classroom experiments are increasingly attracting attention in economics education \citep{holt1999teaching,picault2019economics,platz2022learning}. An example is the paper of \cite{dobrescu2015learning}, assessing and discussing a video game in a standard Microeconomics introductory undergraduate course. Another example is the \textit{Chair the Fed} online simulation. This video game, developed by the Federal Reserve Bank of San Francisco and discussed by, e.g., \cite{duzhak2021effects}, teaches on the Fed's dual-mandate goals of stable prices and low unemployment.  A closely related study to our proposal is presented by \cite{correa2016role}, which employs a role-play scenario to teach the strategic behavior of a cartel, with participants acting as member countries of the Organization of Petroleum Exporting Countries (OPEC). Additionally, \citet{picault2015introduction} introduces a classroom experiment to assess strategic interactions and collusion among producers in a market.

	\section{Classroom game}\label{Sec:Classgame}
	
	The classroom game is divided into three parts. Participants act as construction firms submitting bids to a procurement agency in the first two parts. In Part 1, they act as an oligopoly, with no coordination allowed. That is, they submit bids based on the known distribution of costs as well as on the known value of their own cost. In Part 2, this procedure is repeated, with the difference that explicit coordination among each other is now allowed. In Part 3, the participants switch sides and apply their data science and economics knowledge to flag conspiracies by taking on the role of employees of an antitrust authority. In the following, we will use the term "firms" to denote participants in Parts 1 and 2 and "employees" to refer to them in Part 3. Other roles of the classroom game, such as the procurement agencies and the antitrust authority's board of directors, remain unallocated to participants. Instead, these roles adhere to the game's rule set, overseen by the lecturer. Figure \ref{Parts_Experiment} illustrates the classroom game's action. In Appendix \ref{Appendix:Groupsize}, we suggest a suitable groups allocations depending on the class size.  
	
	\begin{figure}[H] 
		\includegraphics[scale=.38]{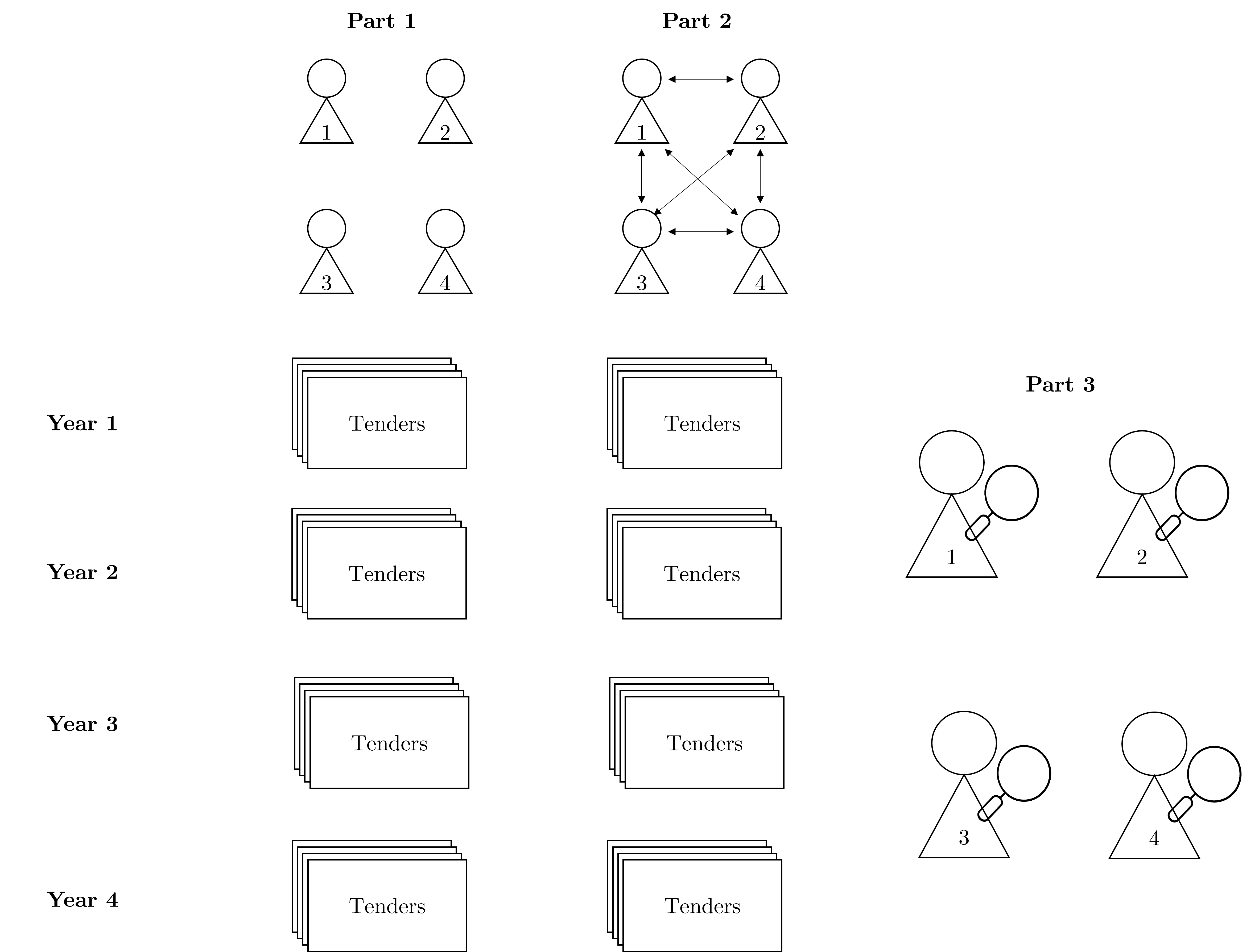}
		\centering 
		\vspace{15pt} 
		\caption{Parts of the classroom game}
		\label{Parts_Experiment}
	\end{figure}
	
	In the following, we first explain the setting in greater detail. Second, we show how firms calculate their costs for a single project (being relevant in Parts 1 and 2). Next, we explain the bidding and the awarding procedure of Parts 1 and 2. Fourth, we explain the machine learning application to screen for cartels in greater detail (Part 3). Finally, we present a concluding example. 
	
	\subsection{Setting} \label{sec:setting}
	
\begin{figure}[H] 
		\includegraphics[scale=0.6]{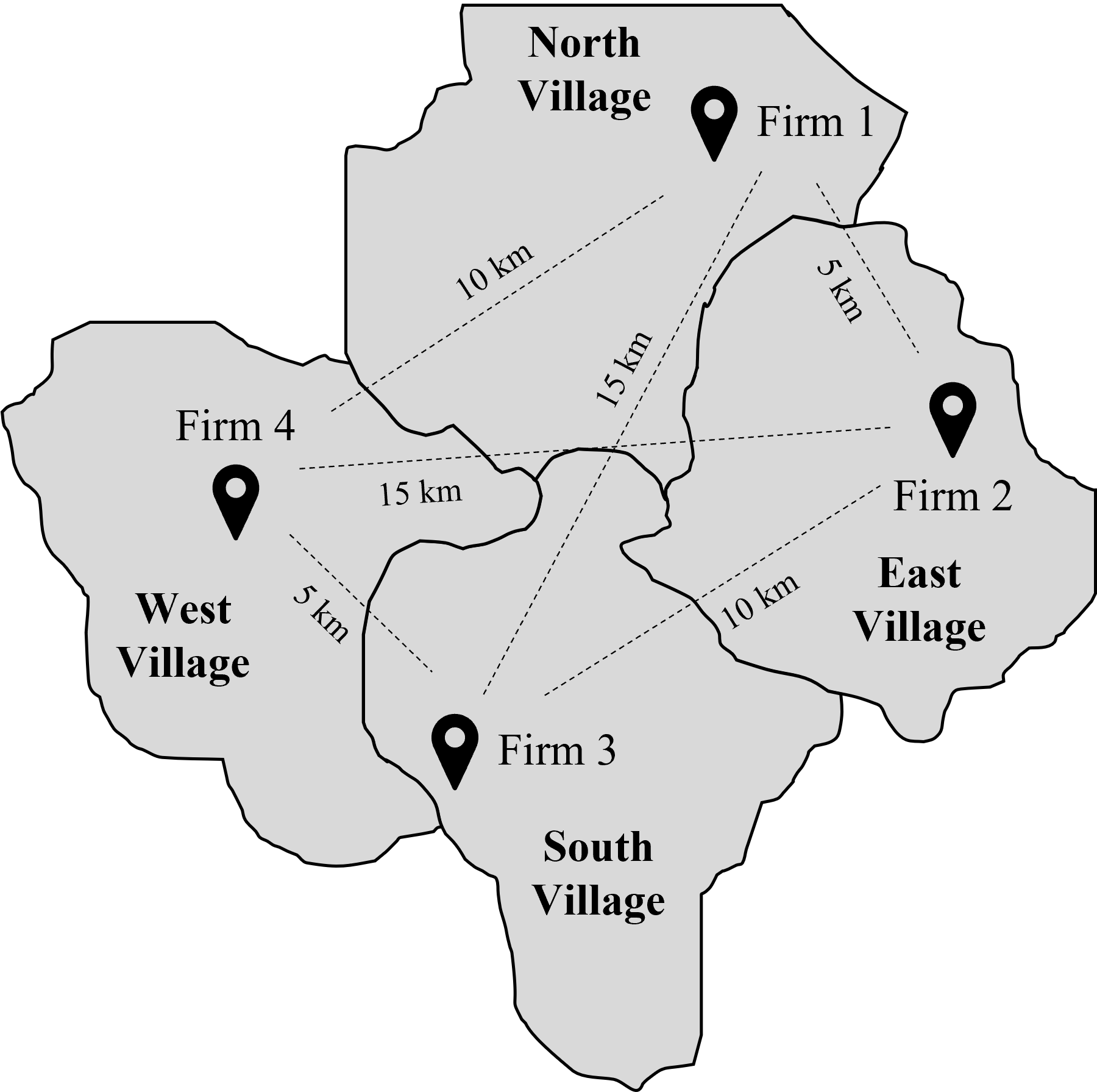}
		\centering \vspace{15pt}
		\caption{The fictional region of Tetravale} \label{Villages}
	\end{figure}

	Revisiting Figure \ref{Parts_Experiment}, we observe that Parts 1 and 2 each contain four years, and each year contains four projects put to tender by a procurement agency. The mentioned procurement agency pertains to the fictional region of "Tetravale" (see Figure \ref{Villages}). Each year, the project locations of the four tenders are "North Village", "East Village", "South Village", and "West Village". All participants are divided into groups of three or four---represented by the four game pieces in Figure \ref{Parts_Experiment}. Accordingly, firms apply for all projects at every location. 
	
	\subsection{Calculation of the costs (Parts 1 and 2)} \label{sec:Calculation}
	
	For each project, firms know their costs while bidding. The first cost component is "fixed costs", which range from 50 to 100. Fixed costs are different for each tender but the same for all firms. However, three superimposed factors also influence the firms' costs. 
	
	The first factor is the geographical location of the firm. As the distance between a firm's location and the place of the planned construction site increases, the firm's cost also increases. As an example, consider the procurement of a construction site in North Village. East Village's Firm 2 is located 5\,km apart (cp. Figure \ref{Villages}). Accordingly, we assume that Firm 2's cost to fulfill the contract in question increases by 5 percentage points of the fixed costs, as we record in Table \ref{Costincrease}. In comparison, South Village's Firm 3 is located 15\,km from the construction site, which increases its total costs by a hefty 15 percentage points of the fixed costs. The upper third of Table \ref{Costincrease} shows all cost increases owing to the distance between a firm's and the construction site's location. 
	
	\begin{table}[]
		\begin{center}
			\caption{Cost increase in percent relative to the fix costs} \label{Costincrease}
			\begin{tabular}{lcccc}
				\hline 
				& \shortstack{\vspace{2pt}\\ \textbf{Firm 1}\\ (North \\ Village)} & \shortstack{\textbf{Firm 2}\\ (East \\  Village)} & \shortstack{\textbf{Firm 3}\\ (South \\ Village)} & \shortstack{\textbf{Firm 4}\\ (West \\ Village)} \\ \hline
				\shortstack{\vspace{2pt}\\\textbf{Tender in}}             &                 &                 &                 &                 \\
				North Village                  & 0               & 5               & 15              & 10              \\
				East Village                   & 5               & 0               & 10              & 15              \\
				South Village                  & 15              & 10              & 0               & 5               \\
				West Village \vspace{5pt}                  & 10              & 15              & 5               & 0                \\
				\textbf{Capacity utilization}                 &                 &                 &                 &                 \\
				Situation A                        & 0               & 1               & 2               & 3               \\
				Situation B                        & 3               & 0               & 1               & 2               \\
				Situation C                        & 2               & 3               & 0               & 1               \\
				Situation D \vspace{5pt}         & 1               & 2               & 3               & 0                \\ 
				\textbf{Type of contract}      &                 &                 &                 &                 \\
				Road construction              & 0               & 2               & 4               & 6               \\
				Railway track construction     & 2               & 0               & 6               & 4               \\
				Bus station construction       & 4               & 6               & 0               & 2               \\
				Civil engineering 			   & 6               & 4               & 2               & 0               \\ \hline
			\end{tabular}
			\par 	{\small \smallskip }
			\textit{Note: Each year contains four tenders, i.e., four rounds.}
		\end{center}
	\end{table}
	
	The second cost-driving factor is capacity utilization, which can be influenced by factors such as projects from private clients. Within our classroom game, we simulate dynamic fluctuations in capacity utilization across rounds. Firms experiencing high capacity utilization, e.g., Firm 4 in Situation A (see Table \ref{Costincrease}), can only implement the project at increased costs. The rationale behind this is twofold. First, working at full capacity necessitates the temporary hiring of additional personnel, resulting in administrative obligations. Second, the cost increase can also be attributed to implicit expenses (opportunity costs). In certain scenarios, the assumption of a contract may lead to postponing potentially lucrative projects.
	
	In reality, firms' bidding behavior is influenced by many criteria. These encompass non-monetary elements such as the inclination to engage in projects with societal and environmental aspirations. Additionally, there may be long-term considerations, such as evaluating a project's impact on the firm's portfolio and reputation. For simplicity, we distill these multifaceted considerations into simple proxies in the shape of additional firm-specific cost increases. Consistent with existing literature \citep[see, e.g., ][]{wallimann2022machine}, we assume comparatively small differentials in the firms' evaluations of these factors. To illustrate this point in the classroom game, we introduce a third influencing factor. We assume the existence of four distinct project types: "road construction", "railway track construction", "bus station construction", and "civil engineering". Each firm is specialized or at least inclined towards the execution of a specific type. As exemplified by Table \ref{Costincrease}, when a road-construction project is put up for tender, the costs of Firms 1, 2, 3, and 4 experience increases of 0, 2, 4, and 6 percentage points, respectively, over their fixed costs. 
	
	To summarize, let us consider an example regarding Firm 1, situated in North Village: In Round 1, the procurement agency of West Village is tendering for the construction of new tracks (railway track construction), Firm 1's capacity utilization is low (Situation A), and all firms' fixed costs are 100. The costs of Firm 1 amount to $100\times(1+0.1+0+0.02)=112$. Next, in Round 2, the procurement agency of East Village puts a civil-engineering project to tender, Firm 1's capacity utilization is higher (Situation $C$), and all firms' fixed costs are 75. In this case, the costs of Firm 1 amount to $75\times(1+0.05+0.02+0.06)=84.75$ (see Table \ref{Costincrease}).
	
	As implied by this example, the combination of a project's location, a firm's capacity utilization, and the type of the contract are introduced with an element of randomness. Throughout the progression of the four rounds, competitiveness is balanced across all participating firms (whether there are three or four firms associated with the tender, confer to  Table \ref{tab:groupsize} Appendix \ref{Appendix:Groupsize}). Note, however, that instances may arise where a firm's competitive advantages or disadvantages concentrate within specific scenarios.
	
	\subsection{Bidding and awarding procedure (Parts 1 and 2)} \label{sec:bidding}
	
	Each year, the firms simultaneously receive tenders for four projects. They prepare and submit their bids, drawing upon their self-calculated cost structures as well as their expectations of competitors' actions. Enabling a dynamic, albeit tacit, interaction, the firms receive information about the allocation of contracts, as detailed out below. This process repeats four times in both Part 1 and Part 2. In Part 1 of the classroom game, the participants are not allowed to communicate with each other, while in Part 2, they get an opportunity to talk to each other before bidding. However, they are advised to be careful when doing so: The digital trail they leave behind could raise red flags with the antitrust authority. Without going into specifics, the firms are informed that such circumstances could necessitate the reimbursement of profits derived from the implicated tender. 
	
	Following each year (encompassing each four tenders) in Parts 1 and 2, the procurement agencies allocate contracts to firms with the lowest bids and inform them accordingly. All participants are informed about the winner (player) but not the winner's bid. Additionally, the winner is notified on the profit margin, computed as the difference between his or her cost and bid. This margin is directly translated to victory points.  Firms that do not secure contracts receive zero victory points. Note that in the tenders of Part 2, the awarding of victory points is of a temporary nature, as detailed below. The point-assignment process applies across all 32 tenders within each group (four tenders per year and four years per part over the course of two parts).
	
	\subsection{Flagging cartels (Part 3)}\label{flagging}
	In the final part of the classroom game, participants take on the role of staff members within the antitrust authority tasked with overseeing Tetravale's markets. They comprehensively evaluate the pooled tenders presented in Parts 1 and 2, scrutinizing them for potential cartel activities. Then, they advise their board of directors to decide which cases merit more in-depth investigations. As we explain below, successful advice leads to recognition, later converted into victory points. 
	
	To clarify the awarding process, note that after the screening of tenders (which can be done as a homework assignment in more advanced classes or as an in-class activity in less detailed courses), each employee classifies each tender as either “suspicious” or “non-suspicious”. To do so, the antitrust authority arranges the tenders in a random way. By applying code as provided in the example in Appendix \ref{Appendix_Code}, the employees are then able to prepare a data set with every tender as an observation containing an ID variable and the screens as predictors in separate columns of the data set. As a result of the rearrangement of the tenders, employees can no longer (or, at least, hardly) recognize their own tenders from Parts 1 (actual value non-suspicious) and 2 (actual value suspicious). Next, they, can develop their machine-learning-based codes. To this end, following \cite{wallimannsticher2023tracks}, the antitrust authority provides the employees with data from the three cartels from Switzerland See-Gaster, Grisons, and Ticino, previously also introduced by \cite{imhof2018screening} and \cite{wallimann2022machine}. 
	
	Based on this data, the participants estimate their models. Depending on their level of data-science knowledge, prepared code can be distributed or participants may the freedom to choose algorithms and code (see Section \ref{Sec:Recomm}). In the baseline proposal of our classroom experiment, we suggest an analysis using ready-made "screens". Screens are descriptive statistics of bids describing bidders' behavior in a tender. As predictors for our binary outcome variable---which takes on the value 1 when firms collude and the value 0 when they compete---, we use the coefficient of variation (CV), the spread (SPD), the difference between the two lowest bids (DIFFP), the relative distance (RD), and the normalized distance (NORMRD). We formally introduce the screens in Appendix \ref{Sec:Formal}. Applying such screens, participants should be enabled to capture distributional changes due to a cartel's coordination of bids. Having developed their models, the participants now classify the tenders of Parts 1 and 2 as suspicious or non-suspicious. 
	
	The incentives for the participants are as follows: In Part 3, no additional victory points can be accrued. Instead, the points obtained in Parts 1 and 2 are considered and modified to determine the winner over the course of the three parts. As members of the antitrust authority, the participants' primary goal is to accurately classify as many tenders as possible. Therefore, to determine the winner of the competition, only participants in the top half of Part 3 are considered. Specifically, to be eligible to win, a participant's "correct-classification rate" (comprising true-positive and true-negative classifications) must meet or surpass the median of the correct-classification rate of the class. 
	
	Furthermore, to address the significance of false-positive accusations, which the antitrust authority aims to minimize for reputational reasons (also in light of the heavy legal consequences of a firm being flagged as a potential cartel participant), each incorrect labeling of a "suspicious tender" results in a 5\% reduction of the a participant's victory points. 
	
	Before this deduction takes place, it is important to note that the points awarded in Part 2 are provisional: Once at least 75\% of the antitrust authority's staff members classify a tender as suspicous, it is deemed "unequivocal suspicious". This initiates a more extensive investigation, including key-witness interviews and raids, to provide a vivid illustration. In cases where a tender was actually a cartel (i.e., a tender of Part 2), the victory points provisionally awarded in Part 2 are revoked. 
	
	
	\subsection{Concluding example}
	
	The classroom game concludes with the unveiling of the victory point progression. As an illustration, consider a scenario involving a class composed of twelve participants, leading to the formation of three groups, each comprising four firms. Focus on Participant $A$'s firm and its markups during the four years of competition and collusion, respectively:
	\begin{align*}
		\text{Competition (Part 1): }& \{\overbrace{(0,0,0,0)}^{\shortstack{\scriptsize Part 1, \\ \scriptsize Year 1}}, \overbrace{(0, 0, 0,0)}^{\shortstack{\scriptsize Part 1, \\ \scriptsize Year 2}}, \overbrace{(10, 10, 0,0)}^{\shortstack{\scriptsize Part 1, \\ \scriptsize Year 3}}, \overbrace{(0,0,0,3.5)}^{\shortstack{\scriptsize Part 1, \\ \scriptsize Year 4}}\}, \\
		\text{Collusion (Part 2): }& \{\underbrace{(0,0,10,0)}_{\shortstack{\scriptsize Part 2, \\ \scriptsize Year 1}}, \underbrace{(0, 0, 10,0)}_{\shortstack{\scriptsize Part 2, \\ \scriptsize Year 2}}, \underbrace{(0, 20, 0,0)}_{\shortstack{\scriptsize Part 2, \\ \scriptsize Year 3}}, \underbrace{(0,20,0,0)}_{\shortstack{\scriptsize Part 2, \\ \scriptsize Year 4}}\}.
	\end{align*}	 
	
	In Part 1, Participant $A$ accumulates 23.5 victory points. This includes two markups of size 10 in Year 3 and one markup of 3.5 in Year 4. Likewise, $A$ secures 60 victory points from Part 2, but it is essential to understand that these points are provisional, given the impending investigation in Part 3. 
	
	Switching sides in Part 3, all participants submit their lists of suspicious and non-suspicious tenders. The antitrust authority subsequently scrutinizes all tenders classified as suspicious by at least 75\% of the participants. In this context, assume that all tenders within Years 3 and 4 of Participant $A$'s group fall within this category of unequivocal suspicious tenders. Consequently, $A$ forfeits 40 of the 60 victory points gained in Part 2. 
	
	Before considering her performance as a staff member of the antitrust authority, $A$'s vicotry points total $23.5+(60-40)=43.5$.

	Now, assume that $A$ also ranks among the top half of the class in terms of the correct-classification rate in Round 3. Furthermore, suppose that she mistakenly labels eight competitive tenders from Round 1 as suspicious. In this case, $A$ remains eligible to win but experiences a 40\% reduction of her victory points, corresponding to the eight incorrect classifications at 5\% each.
	
	In total, across Parts 1 to 3, Participant $A$ accrues $43.5\times 60\%=26.1$ victory points, positioning her as a contender for the main prize, as predetermined by the lecturer. We recommend considering a symbolic prize, perhaps a luncheon voucher, to underscore the intrinsic motivation associated with winning the game.
	
	\section{Didactic recommendations}\label{Sec:Recomm}
	
	The integration of machine-learning techniques will become increasingly important in social sciences in the future. The research on cartel screening serves as a typical example of this trend. This shift in research methods presents challenges for both students and practitioners, as individuals may excel in either economics or data science but less frequently in both.  Our classroom game serves as an example of how social-science educators can incorporate the interdisciplinary nature of modern economics into the classroom without overstraining the participants' capacities. More generally, it demonstrates how scientific articles can be taught in an application-oriented manner.

	In the following, we discuss our findings from an initial in-class realization. Thereupon, we summarize considerations for implementation, emphasizing different settings, participant compositions, and states of knowledge, respectively. The appendices in this paper also provide further basic information on how teaching materials can be designed.

	\subsection{Insights from an initial implementation}\label{Subsec:Insights}
	
	\textit{The working-paper version of this article will be published before an initial implementation will have taken place.}
	
	\textit{Note that the classroom game will be provided as an interactive tool in the .xlsm format (Microsoft Excel documents with macros). This tool facilitates interactions between participants and the lecturer and automates computations associated with "non-playable characters," such as procurement agencies and the antitrust authority. Please note that these files, along with related instructions, are not published in connection with the working-paper version of this article. Extensive playtesting in class must occur first.}
		
	\subsection{Recommendations based on data-science knowledge}\label{Subsec:Settings}	
	
	One challenge of embedding data science in (economics) lessons is the question of the participants' knowledge of data science and machine learning. We categorize participants with respect to their knowledge according to three groups: 
	\begin{enumerate}[\itshape i.]
	\setlength\itemsep{0pt}
		\item Participants with advanced data-science knowledge, 
		\item participants with basic data-science knowledge, 
		\item participants with no (or hardly any) data-science knowledge. 
	\end{enumerate}

	Participants from the advanced group can train models themselves using algorithms and statistical markers of their choice. In instances with even better trained participants, they could even apply unsupervised machine-learning algorithms \citep[see, e.g., ][]{silveira2023you}. 
	
	In the second case, i.e., when preexisting knowledge is limited to the basics, it is advisable to specify specific algorithm-training processes. As an example, the code in Appendix \ref{Appendix_Code} demonstrates the application of the random-forest algorithm. The utilization of predefined code has the advantage that the analysis can be carried out in the classroom even when time is limited. However, this approach also has two disadvantages: First, the uncritical adoption of existing code contributes less to the understanding of the subject. Second, all participants will obtain identical results. (There are only deviations if participants use different "random seeds"---or if they make mistakes).  To alleviate these issues, participants can be encouraged to modify the critical value for the variable "cartel probabilities" (variable "threshold" in the code of Appendix \ref{Appendix_Code}). This adjustment influences the random-forest algorithm's classification of a bid, indicating the presence of a cartel. In doing so, participants who recognize that "false positives" can have particularly severe consequences in the cartel-flagging process (see Subsection \ref{flagging}) have the opportunity to distinguish themselves by classifying potentially collusive bids more conservatively. As this choice entails the risk of not achieving a sufficiently high "correct-classification" rate, the additional degree of freedom introduces an engaging new tradeoff in addition to gaining a deeper understanding.
		
	Finally, in the third case, i.e., when the group has little or no data-science knowledge, a practical approach involves visual or benchmark-based screening \citep[see, e.g., ][]{imhof2019detecting}. For instance, empirical evidence suggests that colluding firms often submit bids that are closely aligned \citep[see, e.g., ][]{imhof2018screening}. While more advanced groups incorporate this insight by integrating the statistical markers into the algorithm's training, the same pattern can be discerned through visual representations of bid distributions (utilizing scatterplots) in widely available statistical software like Microsoft Excel. Also when examining other markers introduced in Appendix \ref{Sec:Formal}, conspicuous values can be identified visually: Low values of the coefficient of variation (CV) link to collusion. The spread (SPD) assumes high values when the lowest and highest bids diverge substantially. DIFFP represents the difference between the two lowest bids. The relative distance (RD) and normalized relative distance (NORMRD) scale the difference between the two lowest bids relative to the variation of the losing bids and the differences between any bids, respectively. In class, the lecturer can either present scientific evidence on these markers or engage students in a discussion about patterns that may raise suspicion. The subsequent classification within needs to necessarily involve graphical representations of the bid distribution. Participants could also classify tenders based solely on the numerical values of the bids.

	Being best equipped to assess the proficiency of their classes themselves, lecturers have the flexibility to tailor our suggestions to meet the unique needs of the participants. Even in classes with advanced data-science knowledge, incorporating visual analysis may prove worthwhile. It not only acknowledges and rewards participants for their intuitive understanding of the subject, which may capture nuances in the bids distribution "overlooked" by an algorithm. The possibility of overwriting machine-learning-based classifications also introduces greater heterogeneity in the cartel-flagging process, fostering additional game strategies and thereby adding excitement.	
	
	Be aware that introducing the setting, especially in Part 3, may be time-consuming. While Parts 1 and 2 of the game are intended for in-class play, Part 3 can be structured as homework assignments. This allows for a brief discussion of Part 3 during the following class (alongside awarding the winner). Furthermore, this assignment has the potential to enhance participants' competence acquisition, given that the task can be completed at individual paces.

	\section{Conclusion}\label{Conc}
	
	In this study, we showcased how game-based learning can be deployed to address the challenges arising from the multidisciplinary nature of modern economics. To do so, we presented a classroom game that integrates economic theory with the application of machine learning.  The first two parts involve participants acting as firms in an oligopoly, submitting bids for construction projects. Communication is restricted in Part 1 and enabled in Part 2. In the third part, participants assume the role of employees in a competition authority, pursuing the goal of detecting illegal collusion in bid patterns. The classroom game's design, breaking it into three parts and combining it with theoretical inputs, addresses concerns about potential short-lived engagement raised by critics of classroom games. This approach supports the contribution of the learning objectives in that participants remain motivated throughout the entire learning experience. Moreover, a focus of our paper lies within Section \ref{Sec:Recomm}, where we provide educational suggestions inter alia for groups with different data science knowledge. Moreover, in the Appendix, we added teaching material tailored to the classroom game.
	
	Our classroom game serves as a practical and engaging approach to address the evolving landscape of cartel detection in the context of economics and data-science education. Through the three parts of our game, participants not only gain insights into the dynamics of collusion in markets, illustrated by the construction sector, but also develop skills in utilizing data-science techniques for cartel detection by examining bid data for suspicious patterns. As the use of machine learning in cartel screening becomes increasingly prominent in industrial organization---both in academics and in the professional world---our classroom game provides a platform for students and practitioners to acquire relevant skills in this evolving field. Incorporating this classroom game into (multidisciplinary) economics education can contribute to preparing the next generation of professionals to be better equipped to address the challenges posed by cartels in the era of big data.
	
	\newpage
	\bigskip
	
	\bibliographystyle{econometrica}
	\bibliography{EconomicsInAction.bib}
	
	\pagebreak
	
	\begin{appendix}
		
		\numberwithin{equation}{section}
		\counterwithin{figure}{section}
		\counterwithin{table}{section}
		\noindent \textbf{\LARGE Appendices}

		\section{Recommended group sizes} \label{Appendix:Groupsize}
		For practical purposes, we suggest suitable allocations into groups of 4 and 3 for any class size from 6 to 32. (With even more extensive classes, due to handling considerations, we recommend splitting up the group into two half classes and playing the game in two separate sessions. With less than 6 participants, all players could be part of one group.) The rule behind the recommendations recorded in Table \ref{tab:groupsize} is to form as many groups of size four as possible while being able to allocate the remaining participants into groups of 3. The asymmetry, which results in most situations, leads to a slightly unequal playing field: Specifically, it is easier to succeed in a tender when only three instead of four firms are involved. However, as the groups are put together randomly, this bias only occurs ex-post. Furthermore, outside of sports, there are barely any games that do not involve luck, at least to some extent. (Chess as a so-called "perfect information" game is one of the rare exceptions.) Accordingly, there is no need to discourage lecturers from playing with unequally-sized groups.
		
		\begin{table}[H]
			\caption{Group recommendations as a function of class size $N$ (preceding the colon)} \label{tab:groupsize}
			\centering
			\resizebox{\columnwidth}{!}{%
				\begin{tabular}{lllll}
					\hline
					6: $2\times 3$ participants                 &  & 15:  $3\times 4$  + $1\times 3$ participants &  & 24: $6\times 4$ participants                \\
					7: $1\times 4$  + $1\times 3$ participants  &  & 16: $4\times 4$ participants                 &  & 25: $4\times 4$  + $3\times 3$ participants \\
					8: $2\times 4$ participants                 &  & 17: $2\times 4$ + $3\times 3$ participants   &  & 26: $5\times 4$  + $2\times 3$ participants \\
					9: $3\times 3$ participants                 &  & 18: $3\times 4$ + $2\times 3$ participants   &  & 27: $6\times 4$  + $1\times 3$ participants \\
					10: $1\times 4$  + $2\times 3$ participants &  & 19: $4\times 4$  + $1\times 3$ participants  &  & 28: $7\times 4$ participants                \\
					11: $2\times 4$  + $1\times 3$ participants &  & 20: $5\times 4$ participants                 &  & 29: $5\times 4$  + $3\times 3$ participants \\
					12: $3\times 4$ participants                &  & 21: $3\times 4$  + $3\times 3$ participants  &  & 30: $6\times 4$  + $2\times 3$ participants \\
					13: $1\times 4$  + $3\times 3$ participants &  & 22: $4\times 4$  + $2\times 3$ participants  &  & 31: $7\times 4$  + $1\times 3$ participants \\
					14: $2\times 4$  + $2\times 3$ participants &  & 23: $5\times 4$  + $1\times 3$ participants  &  & 32: $8 \times 4$ participants \\ \hline             
				\end{tabular} 
			} 
			
		\end{table}

		\section{Formal introduction of the screens}\label{Sec:Formal}
		
		Following \cite{wallimannsticher2023tracks}, we formally introduce the screens we use as predictor variables $X$. $t$ denotes a tender $t$; ${\bar{b}_{t}}$ denotes the mean of all bids in tender $t$; $sd_{t}$ and $sd_{\text{losing bids},t}$ represent the standard deviations of all bids and the losing bids (not including the lowest one of tender $t$) respectively; $n_{t}$ denotes the number of bids submitted in tender $t$; $b_{\max,t}$ is the highest bid of tender $t$; $b_{i,t}$ is the $i$th bid of tender $t$, ordered from lowest to highest bid. To calculate the screens we apply the following formulas: 
		
		\begin{tabular*}{\textwidth}[b]{p{0.5\textwidth}p{0.4\textwidth}}
			\begin{equation}\label{eqcvMLS}
				\text{CV}_{t}=\frac{sd_{t}}{\bar{b}_{t}},
			\end{equation}
			& \begin{equation} \label{SPD}
				\text{SPD}_{t}=\frac{b_{\max,t}-b_{1,t}}{b_{1,t}},
			\end{equation}
			\\[-15pt]
			\begin{equation} \label{DiffPerMLS}
				\text{DIFFP}_{t}=\frac{b_{2,t}-b_{1,t}}{b_{1,t}},
			\end{equation} 
			& \begin{equation} \label{RDTMLS}
				\text{RD}_{t}=\frac{b_{2,t}-b_{1,t}}{sd_{\text{losing bids},t}},
			\end{equation} 
			\\[-35pt]
			
			\begin{equation} \label{RDNORMTMLS}
				\text{RDNOR}_{t} = \frac{b_{2,t}-b_{1,t}}{\frac{1}{n_{t}-1}\sum_{i=1}^{n_{t}-1}(b_{i+1,t}-b_{i,t})}.
			\end{equation} & 
		\end{tabular*}

		\section{Code example}\label{Appendix_Code}
				
		The code for the statistical software \textsf{R} provided below builds upon the output of the tool that leads through Parts 1 and 2 of the classroom game.
		
		In "Load and prepare data", we install and load the \textit{readxl}, \textit{dplyr}, \textit{randomForest}, and \textit{tidyverse} packages for \textsf{R} using the \textit{install.packages} and \textit{library} commands. Then, we load into our \textsf{R} workspace the data of Parts 1 and 2.
		
		In "Prepare screens", we prepare the data by computing the screens discussed in Subsection \ref{Subsec:Settings}). We propose this part of the code also being provided by the lecturer to participants of any knowledge level, as programming it in class is time consuming and does not substantially contribute to the learning objective of understanding how data science is employed to detect cartel behavior. Specifically, we prepare the screens (SPD, CV, RD, RDNORM, DIFFP) we use for screening. In conjunction with the data provided by the classroom-game tool, we obtain a dataset containing anonymized observations including screens, bids, and an identifier.
		
		In "Add the bids to the sample", we rank the bids of each tender and add them to the observation, which now includes the tender (defined by group, year, and round), screens, bids, and an identifier.
		
		We conclude the data preparation in "Finalize and export data set" before moving on to the algorithm training. (Thanks to \cite{rodriguez2022collusion}, data on such cartels are publicly available: \url{https://www.sciencedirect.com/science/article/pii/S0926580521004982} (accessed on January 26, 2024).) Moreover, lecturers without access to a suitable dataset might consider other methods for screening, such as described in Section \ref{Sec:Recomm}).
			
		In "Prepare screening", we load into our \textsf{R} workspace the data of the Ticino, See Gaster, and Grison cartels, where the cartels' incidence is known \citep[see, e.g., ][]{wallimannsticher2023tracks}. Using the \textit{set.seed} command, we make sure that we get the same results while rerunning the code. Then, we split our data into training and test sets. 
			
		In "Train model", by calling the \textit{randomForest} command, we first investigate our model's performances. To do so, we predict the presence of a cartel in the test set by using \$ to call and create subobjects in \textsf{R}. The thing to notice is that we make probability predictions in order to set the probability threshold. Then, by comparing the mean difference between the actual and predicted outcomes, we can calculate the correct classification rate (accuracy). Running this \textsf{R} code yields a accuracy of 0.834. Put simply, we correctly classify 83.4\% of the cases in the test set. (Note that other evaluation metrics, e.g., the F1 score, would be appropriate in cases with non-balanced data sets. For simplification, we present the accuracy here.)	
		
		In "Classifications in Part 3", we apply the trained model ("rf") the prepared data to classify the tenders. Note that we, again, apply a probability threshold of 0.5. 

		Finally, in "Write CSV for submission to lecturer", we conclude the analysis and export a csv document that can be submitted to the lecturer. \\	
		
		\footnotesize\noindent\texttt{\# Load and prepare data}\\
		\footnotesize\noindent\texttt{install.packages("readxl"); install.packages("dplyr"); install.packages("tidyverse")}\\
		\footnotesize\noindent\texttt{install.packages("randomForest")}\\
\footnotesize\texttt{library(readxl); library(dplyr); library(tidyverse); library(randomForest)}\\
		\footnotesize\texttt{ds <- read\_excel("/path/to/my/directory/Example.xlsx")}\\
		\footnotesize\texttt{ds <- ds |> group\_by(Part, Year, Round, Group) |> mutate(ID = cur\_group\_id())}\\                                      
		\footnotesize\texttt{ds <- ds |>add\_count(ID, name = "nbids")}\\
		
		\footnotesize\noindent\texttt{\# Prepare screens}\\
		\footnotesize\texttt{ds <- ds |>group\_by(ID) |>mutate(max = max(Bid)) \# max}\\
		\footnotesize\texttt{ds <- ds |>group\_by(ID) |>mutate(min = min(Bid)) \# min}\\
		\footnotesize\texttt{ds <- ds |>group\_by(ID) |>mutate(SPD = (max-min)/min) \# SPD}\\
		\footnotesize\texttt{ds <- ds |>group\_by(ID) |>mutate(mean = mean(Bid)) \# mean}\\
		\footnotesize\texttt{ds <- ds |>group\_by(ID) |>mutate(sd = sd(Bid)) \# sd}\\
		\footnotesize\texttt{ds <- ds |>group\_by(ID) |>mutate(CV = sd/mean) \# CV}\\
		\footnotesize\texttt{ds <- ds |>arrange(ID,Bid) |>group\_by(ID) |> mutate(rank = rank(Bid, ties.method = "first")) \#rank }\\
		\footnotesize\texttt{ds <- ds |>group\_by(ID) |>mutate(secondbid = ifelse(rank==2,Bid,NA)) \# second-lowest bid}\\
		\footnotesize\texttt{ds <- ds |>group\_by(ID) |>fill(secondbid, .direction = "updown") \# secondbid}\\
		\footnotesize\texttt{ds <- ds |>group\_by(ID) |>mutate(sd\_lb=sd(Bid[rank>1])) \# sd loosing bids}\\
		\footnotesize\texttt{ds <- ds |>group\_by(ID) |>mutate(RD = (secondbid-min)/sd\_lb) \# RD}\\
		\footnotesize\texttt{ds <- ds |>group\_by(ID) |>mutate(nbids =n())}\\
		\footnotesize\texttt{ds <- ds |>group\_by(ID) |>mutate(RDNORM = (secondbid-min)*(nbids-1)/(max-min)) \# RDNORM}\\
		\footnotesize\texttt{ds <- ds |>group\_by(ID) |>mutate(DIFFP = (secondbid-min)/min) \# DIFFP}\\		
		
		\footnotesize\noindent\texttt{\# Add the bids to the sample}\\
		\footnotesize\noindent\texttt{ds\$Bid\_1 <- ds\$min}\\
		\footnotesize\noindent\texttt{ds\$Bid\_2 <- ds\$secondbid}\\
		\footnotesize\noindent\texttt{ds <- ds |>group\_by(ID) |>mutate(Bid\_3 = ifelse(rank==3,Bid,NA)) \# third-lowest bid}\\
		\footnotesize\noindent\texttt{ds <- ds |>group\_by(ID) |>fill(Bid\_3, .direction = "updown")}\\
		\footnotesize\noindent\texttt{ds <- ds |>group\_by(ID) |>mutate(Bid\_4 = ifelse(rank==4,Bid,NA)) \# fourth-lowest bid}\\
		\footnotesize\noindent\texttt{ds <- ds |>group\_by(ID) |>fill(Bid\_4, .direction = "updown")}\\

		\footnotesize\noindent\texttt{\# Finalize and export data set for Part 3}\\
		\footnotesize\noindent\texttt{Tender <- ds |>filter(rank==1)}\\
		\footnotesize\noindent\texttt{data <- Tender |>select(ID, SPD,CV,RD,RDNORM,DIFFP,Bid\_1,Bid\_2,Bid\_3,Bid\_4)}\\
		\footnotesize\noindent\texttt{write.csv(data, "/path/to/my/directory/data\_prep.csv", row.names=FALSE)}\\		
		
		\noindent\texttt{\# Prepare screening (example for participants)}\\
		\texttt{data <- read.csv("/path/to/my/directory/data.csv", sep=";")}\\
		\texttt{set.seed(123)}\\
		\texttt{train=sample(nrow(data),nrow(data)*0.75)}\\
		\texttt{data.test=data[-train,]}\\
		\texttt{data.train=data[train,]}\\
		
		\noindent\texttt{\# Train model}\\
		\texttt{rf <- randomForest(factor(cartel)\raisebox{-0.9ex}{\~{ }}SPD+CV+RD+RDNORM+DIFFP,data=data.train)}\\
		
		\noindent\texttt{\# Make predictions (i.e., with probability threshold equal 0.5)}\\
		\texttt{data.test\$predicted.prob <- predict(rf,data.test, type = "prob")}\\
		\texttt{data.test\$predicted.prob <- data.test\$predicted.prob[, "1"]}\\
		\texttt{threshold <- 0.5}\\
		\texttt{data.test\$predicted.response <- ifelse(data.test\$predicted.prob >= threshold, 1, 0)}\\

		\noindent\texttt{\# Assess accuracy}\\
		\texttt{accuracy <- mean(data.test\$predicted.response==data.test\$cartel)}\\
		\texttt{accuracy}\\
		
		\noindent\texttt{\# Classifications in Part 3}\\
		\texttt{data\_part3 <- read.csv("/path/to/my/directory/data\_prep.csv", sep=";")}\\
		\texttt{data\_part3\$predicted.prob <- predict(rf,data\_part3, type = "prob")}\\
		\texttt{data\_part3\$predicted.prob <- data\_part3\$predicted.prob[, "1"]}\\
		\texttt{threshold <- 0.5}\\
		\texttt{data\_part3\$predicted.response <- ifelse(data\_part3\$predicted.prob >= threshold, "collude", "compete")}\\
		
		\noindent\texttt{\# Write csv for submission to lecturer}\\
		\noindent\texttt{data\_part3\_final <- data\_part3 |>select(ID, predicted.response)}\\
		\noindent\texttt{write.csv(data\_part3\_final, "/path/to/my/directory/data\_part3.csv", row.names=FALSE)}\\
		
	\end{appendix}
\end{document}